\documentclass[]{interact}

\usepackage{epstopdf}
\usepackage{subfigure}

\usepackage{hyperref}
\usepackage{lineno}
\usepackage{natbib}
\bibpunct[, ]{(}{)}{;}{a}{}{,}
\renewcommand\bibfont{\fontsize{10}{12}\selectfont}

\theoremstyle{plain}

\theoremstyle{definition}

\theoremstyle{remark}

\begin{document}

\articletype{The International Encyclopedia of Geography: People, the Earth, Environment, and Technology}

\title{Human-centered Geospatial Data Science}

\author{
\name{Yuhao Kang\textsuperscript{a}\thanks{CONTACT Yuhao Kang. Email: yuhao.kang@austin.utexas.edu}}
\affil{\textsuperscript{a}GISense Lab, Department of Geography and the Environment, The University of Texas at Austin, TX, United States; yuhao.kang@austin.utexas.edu
}}

\maketitle

\begin{abstract}
This entry provides an overview of Human-centered Geospatial Data Science, highlighting the gaps it aims to bridge, its significance, and its key topics and research.
Geospatial Data Science, which derives geographic knowledge and insights from large volumes of geospatial big data using advanced Geospatial Artificial Intelligence (GeoAI), has been widely used to tackle a wide range of geographic problems.
However, it often overlooks the subjective human experiences that fundamentally influence human-environment interactions, and few strategies have been developed to ensure that these technologies follow ethical guidelines and prioritize human values.
Human-centered Geospatial Data Science advocates for two primary focuses.
First, it advances our understanding of human-environment interactions by leveraging Geospatial Data Science to measure and analyze human subjective experiences at place including emotion, perception, cognition, and creativity. 
Second, it advocates for the development of responsible and ethical Geospatial Data Science methods that protect geoprivacy, enhance fairness and reduce bias, and improve the explainability and transparency of geospatial technologies.
With these two missions, Human-centered Geospatial Data Sciences brings a fresh perspective to develop and utilize geospatial technologies that positively impact society and benefit human well-being and the humanities.
\end{abstract}

\begin{keywords}
Human-centered Geospatial Data Science; place; ethics; geospatial big data; GeoAI
\end{keywords}



\section{Introduction}
The complex relationships and interactions between humans and their natural environment have long been a central topic in geography, a focus that dates back to the Man-Land Tradition \citep{pattison1964four}. 
Geographers have studied the dual relationship between humans and the natural world, examining how humans adapt to, modify, and affect their surroundings, and conversely, how the environment shapes human activities and cultural practices \citep{tuan1990topophilia}. 
The introduction of Geographic Information Systems (GIS) and, more recently, Geospatial Data Science has provided unprecedented opportunities to understand human-environment relationships through advanced data collection methods and enhanced spatial analysis and modeling. 
Geospatial Data Science, the discipline of extracting geographic knowledge from geospatial big data, utilizing advanced geocomputational methods such as Geospatial Artificial Intelligence (GeoAI) \citep{janowicz2020geoai}, could significantly enhance our understanding of human-environmental interactions.
Specifically, the emergence of geospatial big data including large-scale human mobility data, user-generated social media content, extensive street view images, and remote sensing images, offers solid data foundations for digitally representing and understanding the world from a data-driven perspective.
Moreover, advanced geocomputational techniques, ranging from GIS to cutting-edge Geospatial Artificial Intelligence (GeoAI), allows for the effective handleing of complex geographical tasks, such as spatial reasoning, geographical phenomena modeling, and geospatial analysis and simulations. 
Numerous applications utilizing GIS and Geospatial Data Science have been developed, benefiting various domains ranging from large-scale land use monitoring, effective resource management, emergency response and hazard mapping, to smart city planning, etc. 
Despite their success, researchers have observed two challenges within the current GIS and Geospatial Data Science.

First, the current development and applications of GIS and Geospatial Data Science have primarily been driven by technology, and have paid relatively insufficient attention to human experience. 
Geographic phenomena are often treated as functions of physical settings such as terrain, climate, and infrastructure, while neglecting the subjective experiences of individuals and communities. 
Contemporary research in GIS and Geospatial Data Science often adhere to static spatial notions \citep{kwan2013beyond}, concentrating on the tangible properties of space \citep{freundschuh1997human,giordano2018limits}.
This perspective overlooks the dynamic and complex ways in which humans interact with and perceive their environments \citep{shaw2021understanding}, potentially missing the nuanced and varied subjective experiences of different individuals and groups within the physical space.

Humans experience place rather than abstract space.
Beyond space, place is infused with human meanings, encompassing individual social connections, attitudes, and values, all rooted in the subjective nature of human experience.
People learn about and interpret their surroundings by utilizing their senses, such as vision, touch, and hearing, forming what is known as \textit{sense of place} \citep{tuan1977space}. 
These senses guide our behaviors and shape our interactions with our environments, reflecting our \textit{mental space} \citep{shaw2021understanding}.
Consider the experience of coming home after an exhausting day at work.
Your home, located in a neighborhood perceived as safe and welcoming, including familiar streets and friendly neighbors, may evoke positive emotions of comfort and a sense of belonging, fostering meaningful and supportive environments.
What constitutes human subjective experiences such as emotions, perceptions, cognition, and creativity toward various environmental settings?
Leveraging Geospatial Data Science not only helps answer this question but also enables researchers to gain a deeper understanding of the diverse ways people perceive and experience their environments, leading to more inclusive and effective spatial decision making. 
Furthermore, enriching GIS and Geospatial Data Science with human dimensions can benefit the development of these technologies, ensuring they are more closely aligned with human behaviors \citep{zhao2022humanistic}. 

Second, the social and ethical implications of GIS and Geospatial Data Science has received insufficient attention \citep{crampton1995ethics}.
The use and potential misuse of Geospatial Data Science may pose several challenging ethical issues and raise deep concerns related to geoprivacy, bias and fairness, and explainability and transparency \citep{pavlovskaya2018critical,nelson2022accelerating}.
For instance, large-scale human mobility data can reveal sensitive personal information, such as home addresses and frequently visited locations, emphasizing the crucial need to protect this information from misuse and unauthorized access.
Using biased data for training crime prediction models may lead to increased surveillance and discrimination against certain populations, perpetuating cycles of inequality \citep{mayson2018bias}.
Additionally, algorithms used to support decision-making, such as resource allocation, need to be transparent so that stakeholders can understand how decisions are made and trust the outcomes. 
However, limited strategies have been developed to address these social and ethical issues associated with Geospatial Data Science.
Questions arise such as where is the human element in these technologies, and who are the humans behind these technologies? 
Developing ethical GIScience technology involves prioritizing fundamental human values---fairness, transparency, privacy, safety, trust, inclusivity, sustainability, and legal compliance---to ensure technological innovation respects human rights.
Incorporating these human-centered social and ethical considerations into the development of Geospatial Data Science supports more equitable and inclusive design and decision-making, ultimately creating environments that enhance human life and social well-being.

\begin{figure}
    \centering
    \includegraphics[width=\linewidth]{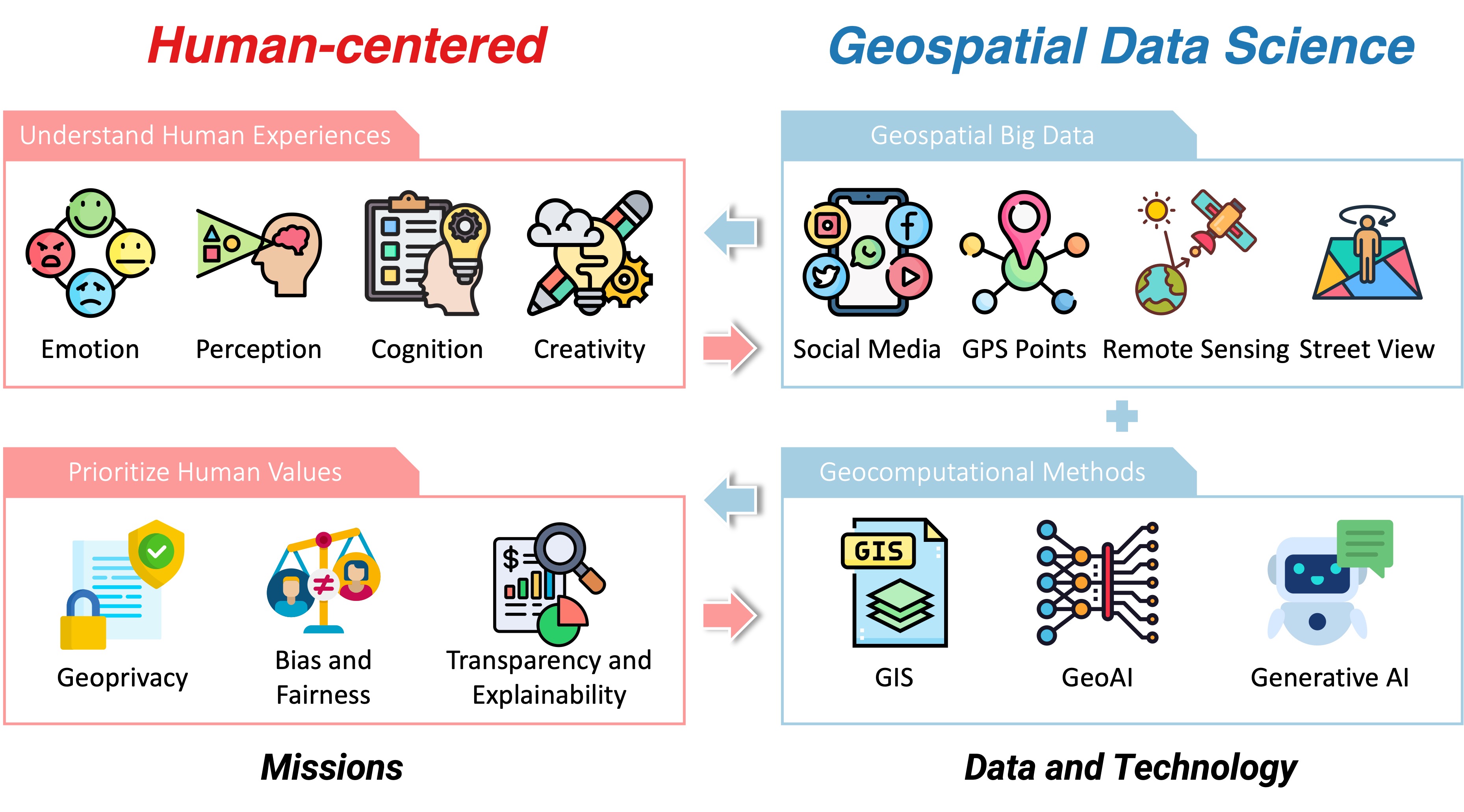}
    \caption{A conceptual framework of Human-centered Geospatial Data Science}
    \label{fig:framework}
\end{figure}

Human-centered Geospatial Data Science aims to understand human experiences and prioritize human values within the field of Geospatial Data Science. 
By adopting a more holistic and people-oriented understanding of geography, this framework bridges critical gaps by focusing on two main missions: (1) advancing human-environment relationships through a deep understanding of human experiences, and (2) developing trustworthy and responsible geospatial technologies that prioritize human values.
Figure \ref{fig:framework} shows an overview of the conceptualization of Human-centered Geospatial Data Science.

\section{Understanding Human Experiences}
Human experiences of place consist of a broad spectrum of internal mental processes, such as emotion, perception, cognition, and creativity.
Traditionally, to collect these human dimensions of experiences, researchers have to rely on questionnaires, surveys, interviews, etc., which are labor-intensive and time-consuming.
Additionally, the inherently subjective nature of these experiences poses significant challenges for accurate measurement and analysis.
Geospatial Data Science, leveraging multi-source geospatial big data and advanced GeoAI methods, offers new opportunities for understanding human subjective experiences. 

\subsection{Emotion}
Different places could evoke a variety of human emotions, deeply influencing our intimate bodily states.
When people interact with various places or events, their emotional responses might be affected by the environmental settings of places \citep{feng2024study}.
For example, a child is likely to feel joy and excitement at Disneyland; while we may feel a sense of solemnity or sadness when visiting a memorial site.
By utilizing Geospatial Data Science, researchers can now map emotional landscapes and analyze how geographical factors influence our emotions.
Geotagged user-generated content has offered new avenues for examining human emotions at places. 
People often express their emotions, both consciously and unconsciously, at various social media platforms. 
Using advanced GeoAI methods, researchers can tap into this rich data source to measure human emotions associated with specific places. 
Prior studies can be primarily categorized into two streams:

First, \textit{text-based analytical methods} utilizes massive geotagged social media textual content and applies Natural Language Processing (NLP) techniques to decipher and analyze the emotions expressed by people. 
Researchers can measure multiple dimensions of human emotions, including happiness, sadness, and positive or negative sentiments, by examining tweets, comments, and posts, 
This provides valuable insights into fluctuations of public sentiment and mood across different geographical spaces.

Another method refers to \textit{image-based analytical methods}.
Using computer vision techniques, researchers could extract and analyze human emotions from facial expressions captured in images. 
From selfies shared on social media to crowd photos at events, this technique allows for the assessment of emotional responses in situ, offering a direct measure of how people emotionally react to the environments.

In addition to these two primary research streams, some emerging psychophysiological technologies, such as functional magnetic resonance imaging (fMRI), electroencephalogram (EEG), electrocardiogram (ECG), electrodermal activity (EDA), and electromyography (EMG) offer new measuring methods for understanding human emotions \citep{zhao2024neural}.
By associating with real-world or virtual environments, researchers could leverage these methods to better model human emotion-environment interactions.

\subsection{Perception}
Human perception of place describes how individuals experience their environment through various sensory dimensions such as sight, sound, and smell. 
The built environment plays a crucial role in shaping human perceptions by affecting visual, auditory, and other experiences \citep{ito2024understanding}.
For instance, a neighborhood with quiet streets, clear sounds of birds, and visually appealing architecture is likely to be perceived as safe and aesthetically pleasing by its residents.
This reflects the complex human-environment interactions, highlighting how environmental factors influence individual perceptions and feelings.
To measure human perceptions of place, two types of geospatial big data have been widely utilized.

First, geotagged user-generated content posted on social media platforms often reflects people's perceptions of specific places or events.
These digital footprints could reveal how individuals experience and interact with various surrounding environments, reflected through the content they share, such as photos, videos, and texts.
Analyzing geotagged social media data can reveal trends and patterns of multiple dimensions of human perceptions, such as human visual, auditory, and olfactory perceptions at different places.

In recent years, street view imagery has provided another valuable data source for assessing human perceptions of the built environment. 
Street view images capture detailed streetscape settings at eye level, providing a realistic perspective of how individuals might perceive their surrounding environment.
Then, researchers leveraged advanced computer vision approaches, such as Convolutional Neural Networks (CNNs) to measure human perceptions of place in response to street view images.
The hypothesis is that people’s subjective perceptions of the built environment captured in street view images can be assessed to reflect their place perceptions.
Researchers have utilized such a computational workflow and measured multiple dimensions of human perceptions of place including safe, beautiful, depressing, and lively.
This method enables researchers to understand the nuanced subjective human experiences, and advance our understanding of human-environment interactions.

\subsection{Cognition}
Human cognition of place reflects the psychological processes of how people conceptualize, interpret, and make sense of geographical environments.
Through synthesizing multi-sensory perceptions of the environment based on past experiences and prior knowledge, individuals interact with their surrounding environments, fostering their place identity, awareness of vague place, and sense of direction and distance.
These cognitive processes could influence key behaviors, such as navigation, planning, and daily decision-making.
Notably, the way in which places are conceptualized can vary significantly among individuals, influenced by the differences in their cultural background, personal experiences, and emotional connections.
These factors may lead to potentially diverse cognitive processes and interpretations, even for the same place, highlighting the challenges in incorporating subjective cognition of place into Geospatial Data Science.
To collect and analyze human cognition of place, multi-sourced social media data has become a valuable data source that reflects how people conceptualize and perceive different environments.
Researchers have leveraged social media data to capture real-time expressions and analyze human mental space, thereby enriching our understanding of place identity, vague cognitive regions, toponyms, and perceived distance and orientation.
For instance, by leveraging crowdsourced social media data, researchers could explore the unique characteristics that contribute to place identity based on associated texts describing places in social media posts.
Researchers have identified and characterized vague cognitive regions and place names---areas without clear physical boundaries but are frequently referenced in the daily lives of individuals. 
Researchers have also examined the sense of distance and sense of orientation to understand how people navigate and make spatial decisions.
Having a deeper understanding of human cognition of place using social media data could benefit various applications in spatial decision-making. We could understand the complete ways how people conceptualize place, leading to potential effective urban planning and personal navigation solutions.

\subsection{Creativity}
Human creativity --- the ability to generate original ideas and expressions --- is another important component of human experience. 
Artists and creators take the inputs from human experiences and adapt them into compelling narratives and innovative solutions \citep{jung2020teaching}.
In Geospatial Data Science and GIS, the outcomes of human creativity are often embodied in the art of map design and geovisualizations, which are used for creating visually pleasing maps and conveying geographic information through visual storytelling \citep{kang2024artificial}.
Traditionally, it has been believed that machines are unable to replicate the artistic and emotional elements of cartographers' creativity and experience due to the inherently subjective nature of artistic expression.
However, advancements in GeoAI methods may revolutionize this field by enabling the modeling of various aspects of map design, such as map styles, visual variables, and map critique. 
Moreover, the recent emergence of Generative Artificial Intelligence (GenAI) has demonstrated exceptional promise in stimulating and enhancing map creativity and aesthetics. 
Researchers have been leveraging text-to-image models such as Stable Diffusion, DALL-E 2, and Midjourney not only to generate maps directly but also to foster innovative and creative ideas that inspire map design.
Additionally, text-to-text models like ChatGPT are utilized to automate the map-making process.
These developments in AI technologies are opening up new possibilities for understanding human subjectivity and creativity, potentially changing the future paradigm of map-making and expanding what we can visualize.

\section{Prioritize Human Values with Ethical Discussions and Practices}
The ethics of GIS and Geospatial Data Science refers to a set of guidelines and principles designed to ensure the alignment of geospatial technologies and methodologies with human values and uphold human rights.
This ethical focus is crucial for guiding the responsible use and development of trustworthy geospatial technologies.
Researchers from diverse academic and professional communities including GIScience, geography, and computer science have provided their perspectives on the ethical discussions in geospatial theories and practices.
However, few strategies have been developed to implement these ethical considerations in geospatial practices \citep{kang2024artificial}.
Despite the complexity of ethical issues, the following subsections outline three main aspects of ethics in geospatial data science: geoprivacy, bias and fairness, and explainability and transparency.
\citep{ilagcrstrand1970people}
Several recent advancements in tackling ethical issues and developing responsible Geospatial Data Science practices are summarized.

\subsection{Geoprivacy}
Human geoprivacy refers to the right and ability of individuals to govern the collection, sharing, and use of their location-based information.
It protects people's locational privacy from unauthorized access and prevents potential misuse of their sensitive geographical data. 
In our daily lives, we frequently interact with devices that are equipped with GPS and make use of location-based services. 
Activities such as sharing geotagged posts on social media platforms, navigating with GPS-enabled devices, and granting mobile apps permission to access our location data can unintentionally disclose our personal information to different entities, including governments and private sectors.
The data that is shared can expose details about our trajectories, activities, and behaviors, emphasizing the crucial need for effective strategies to protect geoprivacy.

To address these concerns, researchers have investigated multiple methods to enhance human geoprivacy in the existing multiple large-scale geospatial datasets.  
For example, data aggregation and geomasking techniques have been widely used to obscure location information to enhance human geoprivacy.
Additionally, advanced GeoAI methods have been employed to further enhance the protection of geoprivacy. 
Generative Adversarial Networks (GANs), for instance, have been utilized to simulate human trajectories and generate synthetic datasets, allowing researchers to analyze patterns without accessing sensitive individual raw data.
Another cutting-edge approach, federated learning, enables the training of AI models on multiple decentralized devices that store local datasets.
This method ensures that the real data remains on the device and only changes to the model are shared across the network, thus protecting individual privacy.
By employing these advanced techniques for protecting geoprivacy, researchers can ensure that geospatial data management is not only responsible and ethical but also cultivating trust to promote the beneficial uses of Geospatial Data Science.

\subsection{Bias and Fairness}
A fundamental principle in Human-centered Geospatial Data Science is recognizing the humans behind technologies and understand how they engage with these systems.
Addressing bias and fairness in Geospatial Data Science is crucial to prevent potential discrimination against any group or individual and to guarantee that outcomes from geospatial technologies are both fair and equitable.
There are two forms of bias commonly seen in current Geospatial Data Science studies.
\textit{Population bias} relates to how demographic characteristics such as age, gender, race and ethnicity, are represented in geospatial data.
Given the individual differences in subjective human experiences and the collection methods employed, datasets frequently become skewed, and favoring certain demographic groups over others.
For example, geospatial data collected from mobile apps typically overrepresent younger urban users, while underrepresenting older rural populations. 
This discrepancy can lead to models that fail to accurately capture the behaviors or needs of the entire population, potentially leading to biased outcomes and discriminatory practices within Geospatial Data Science.
\textit{Spatial bias}, on the other hand, concerns the geographic representativeness within data.
Geospatial data often is not collected uniformly across all regions, leading to discrepancies in coverage that distort the analysis and applications of the data.
For instance, urban areas and developed countries generally have more extensive data coverage. 
While their counterparts may suffer from significant underrepresentation.
This uneven spatial coverage can skew analysis and modeling, resulting in geographic disparities in the accuracy of data and efficiency of geospatial applications.

To enhance fairness and reduce bias in Geospatial Data Science, researchers could focus on both the data collection phase and during the development of algorithms to tackle these challenges.
In the data collection phase, researchers have fused multiple data sources to diversify data sources.
Each data source has its own pros and cons.
Leveraging multi-source geospatial big data may cover a broader spectrum of geographic and demographic characteristics, ensuring a more balanced representation of different areas and populations. 
Moreover, increasing attention has been paid to certain underrepresented and vulnerable population groups and areas, which historically may have been overlooked.
Through specifically targeting these groups, researchers could enhance the diversity and equity of the collected datasets.
Additionally, the success of Public Participatory GIS (PPGIS) has highlighted the benefits of community engagement in geospatial projects.
Given the nature of spatial heterogeneity, Geospatial Data Science need to account for local contexts.
Involving local communities in the data collection process can enrich Geospatial Data Science with their unique insights and local knowledge. 
This collaborative process not only enhances the richness and accuracy of the collected datasets but also fosters greater trust and inclusivity in the data collection process. 
During the algorithm development phase, researchers have explored several bias-mitigation algorithms designed to address and neutralize biases within GeoAI models.
By adjusting weights during the training process, these algorithms may reduce biases and enhance fairness.
These strategies provided possible solutions for addressing bias and promoting fairness in Geospatial Data Science to ensure inclusive practices. 

\subsection{Transparency and Explainability}
The current GeoAI methods utilized in Geospatial Data Science, ranging from machine learning, deep learning, and the emerging Generative AI (GenAI), often face criticism for their ``black-box'' nature.
Thus, the lack of transparency in these models makes it difficult to understand their underlying mechanisms, which in turn hinders human trust.
Enhancing transparency and explainability in Geospatial Data Science involves the ability to decompose, understand, track, and effectively communicate the decision-making processes of GeoAI methods in modeling geographic phenomena.
It is crucial to articulate how these models derive their predictions in ways that users can trust and validate.
This ensures that stakeholders can make informed decisions in a responsible way.
To advance transparency and explainability in Geospatial Data Science, researchers have explored several innovative approaches.
For instance, researchers have placed more emphasis on the development of eXplainable AI (XAI) methods to understand and interpret the complex internal processes of AI models.
These methods could show which variables and how they influence model decisions.
Additionally, researchers have also leveraged advanced visualization strategies to effectively explain the model outputs.
Both strategies could not just deepen our understanding of GeoAI models, but also provide a robust foundation of trust and responsibility of geospatial technologies. 
These efforts ensure that GeoAI models are comprehensible and transparent, fostering their alignment with societal needs and expectations and promoting their responsible use.

\section{Conclusions}
Human-centered Geospatial Data Science is designed to augment human intelligence and benefit society.
Using large-scale geospatial big data and advanced GeoAI, it focuses on understanding how humans interact with the surrounding environments and prioritizing human values in both research and practices of Geospatial Data Science.
By enriching Geospatial Data Science with human subjective experiences, researchers could advance the knowledge of multiple human dimensions including emotion, perception, and cognition of place, and creativity in mapmaking.
Such an approach helps understand the complex human-environment interactions and could inform the development human-centered algorithms.
To build responsible intelligent systems, it is crucial to understand and prioritize human values such as protecting human geoprivacy, reducing bias and enhancing fairness, as well as improving transparency and explainability of geospatial technologies.
These efforts are essential for ethical use and gaining user trust.
Looking beyond, the development of geospatial technologies needs to be guided by human-centric principles to prioritize the needs and well-being of humans.
This will ensure they are not only innovative but also have positive impacts across all sectors of society.

\section*{Cross-references}
See Also: GeoAI and Its Implications, Social Sensing, Data Science, GeoAI and Deep Learning, Platial, Spatial Data Science

\bibliographystyle{apalike}
\bibliography{reference}

\section*{Further Readings}
Thatcher, J., Shears, A., and Eckert, J. (2018). \textit{Thinking Big data in geography: New regimes, new research.} U of Nebraska Press.362

Williams, S., (2022). \textit{Data action: Using data for public good}. MIT Press.

Shneiderman, B., 2022. \textit{Human-centered AI}. Oxford University Press.
\end{document}